\documentclass[12pt]{article}
 \usepackage{epsfig}
 \def\be{\begin{equation}}
 \def\ee{\end{equation}}
 \def\bea{\begin{eqnarray}}
 \def\eea{\end{eqnarray}}
 \usepackage{graphicx}

 \catcode`\@=11
 \def\lsim{\mathrel{\mathpalette\@versim<}}
 \def\gsim{\mathrel{\mathpalette\@versim>}}
 \def\@versim#1#2{\vcenter{\offinterlineskip
 \ialign{$\m@th#1\hfil##\hfil$\crcr#2\crcr\sim\crcr } }}
 \catcode`\@=12

 \catcode`@=12
\begin{document}
 \thispagestyle{empty}
 \begin{flushright}
 UCRHEP-T571\\
 October 2016\
 \end{flushright}
 \vspace{0.6in}
 \begin{center}
 {\LARGE \bf Self-Interacting Dark Matter with\\ Naturally Light Mediator\\}
 \vspace{1.2in}
 {\bf Ernest Ma\\}
 \vspace{0.2in}
 {\sl Department of Physics and Astronomy,\\ 
 University of California, Riverside, California 92521, USA\\}
 \end{center}
 \vspace{1.2in}

\begin{abstract}\
A promising proposal for resolving the cusp-core anomaly in the density 
profile of dwarf galaxies is to allow dark matter to interact with 
itself through a light mediator of mass much less than a GeV.  The 
theoretical challenge is to have a complete renormalizable theory where this 
happens naturally even though dark matter itself may be of the electroweak 
scale, i.e. 100 GeV to 1 TeV.  I propose here such a model, with 
just two neutral complex scalar singlets under a softly broken dark global 
U(1) symmetry.
\end{abstract}

 \newpage
 \baselineskip 24pt
\noindent \underline{\it Introduction}~:\\
The nature of dark matter is an open question.  If it interacts only weakly 
with visible matter, then there are certain astrophysical observations which 
are not consistent with numerical simulations based on this simple 
hypothesis.  One such discrepancy, i.e. that the density profile of dark 
matter in dwarf galaxies is much flatter near the center (core) than 
predicted (cusp)~\cite{dgsfwggkw09}, has prompted the 
idea~\cite{ss00,fkty09,kty16,ktv16} that dark matter interacts with 
itself through a mediator, much lighter than the dark matter itself. 
Many phenomenological studies have been made, 
but the theoretical challenge 
is to understand why the mediator is light, and what other properties it 
may have, all within a complete renormalizable extension of the standard 
model (SM).

In this paper, I propose such a model.  It assumes a global $U(1)_D$ symmetry 
which is softly and spontaneously broken to $(-1)^D$.  It has just two 
neutral complex scalars: $\zeta$ which has $D=1$, and $\eta$ which has 
$D=2$.  The $U(1)_D$ symmetry is broken spontaneously by the vacuum 
expectation value $\langle \eta^0 \rangle = u$.  The dark particles 
are $\zeta_{R,I}$ which have odd $(-1)^D$ and they interact with 
$\eta_R$ which is heavy and $\eta_I$ which is naturally light, because 
it would be massless if $U(1)_D$ is not broken also by an explicit 
dimension-two soft term.  
Now $\eta_{R,I}$ are even under $(-1)^D$.  Whereas $\eta_R$ mixes with the 
SM Higgs boson $h$ at tree level as usual, $\eta_I$ does so only in one loop. 
This radiative mixing is finite and calculable, a phenomenon discovered 
only recently~\cite{m16}.  It is very important because it allows the 
light $\eta_I$ 
to decay quickly to $e^- e^+$ even if its mass is only 35 MeV, thereby 
not disturbing the success of big bang nucleosynthesis in the SM.

\noindent \underline{\it Model}~:\\
Under the assumed $U(1)_D$, the new scalar singlets are
\begin{equation}
\zeta \sim 1, ~~~ \eta \sim 2,
\end{equation}
and all SM particles are trivial.  The scalar potential consisting of 
$\zeta$, $\eta$, and the SM Higgs doublet $\Phi = (\phi^+,\phi^0)$ 
which becomes $(0, v + h/\sqrt{2})$ in the unitarity gauge, is given by
\begin{eqnarray}
V &=& m_0^2 \Phi^\dagger \Phi + m_1^2 \bar{\zeta} \zeta + m_2^2 
\bar{\eta} \eta 
- {1 \over 2} m_3^2 (\zeta \zeta + \bar{\zeta} \bar{\zeta}) 
- {1 \over 2} m_4^2 (\eta \eta + \bar{\eta} \bar{\eta}) 
+ \mu \bar{\eta} \zeta^2 + \mu^* \eta \bar{\zeta}^2  \\ 
&+& {1 \over 2} \lambda_0 (\Phi^\dagger \Phi)^2 
+ {1 \over 2} \lambda_1 (\bar{\zeta} \zeta)^2 
+ {1 \over 2} \lambda_2 (\bar{\eta} \eta)^2 
+ \lambda_{01} (\Phi^\dagger \Phi)(\bar{\zeta} \zeta) 
+ \lambda_{02} (\Phi^\dagger \Phi)(\bar{\eta} \eta) 
+ \lambda_{12} (\bar{\zeta} \zeta)(\bar{\eta} \eta). \nonumber
\end{eqnarray}
Note that $V$ respects $U(1)_D$ in all its dimension-four and dimension-three 
terms, whereas the dimension-two $m^2_3 (m^2_4)$ terms break $U(1)_D$ to 
$Z_2(Z_4)$.  Without the $m^2_{3,4}$ terms, the spontaneous breaking of $V$ by 
$\langle \eta \rangle = u$ would imply that $\eta_I$ is a massless 
Goldstone boson.  Note also that the phases of $\zeta$ and $\eta$ have been 
rotated to render them real, but the trilinear coupling $\mu$ remains 
complex.

Let $\zeta = (\zeta_R + i \zeta_I)/\sqrt{2}$ and 
$\eta = u + (\eta_R + i \eta_I)/\sqrt{2}$, then the minimum of $V$ 
is determined by
\begin{eqnarray}
0 &=& m_0^2 + \lambda_0 v^2 + \lambda_{02} u^2, \\ 
0 &=& m_2^2 - m_4^2 + \lambda_2 u^2 + \lambda_{02} v^2.
\end{eqnarray}
The mass of $\eta_{I}$ is then naturally small because it comes from a soft 
term which breaks $U(1)_D$ explicitly, i.e.
\begin{equation} 
m^2_{\eta_I} = 2 m_4^2.
\end{equation}
The mass-squared matrix spanning $(h,\eta_R)$ is
\begin{equation}
{\cal M}^2_h = \pmatrix{ 2 \lambda_0 v^2 & 2 \lambda_{02} vu \cr 
2 \lambda_{02} vu & 2 \lambda_2 u^2},
\end{equation}
and that spanning $(\zeta_R,\zeta_I)$ is
\begin{eqnarray}
{\cal M}^2_\zeta &=& \pmatrix{m_1^2 + \lambda_{01} v^2 + \lambda_{12} u^2 
+ 2\mu_R u - m_3^2  & -2\mu_I u \cr -2\mu_I u & 
m_1^2 + \lambda_{01} v^2 + \lambda_{12} u^2 - 2\mu_R u + m_3^2} \nonumber \\ 
&=& \pmatrix{\cos \theta & -\sin \theta \cr \sin \theta & \cos \theta} 
\pmatrix{ m^2_{\chi_1} & 0 \cr 0 & m^2_{\chi_2}} 
\pmatrix{\cos \theta & \sin \theta \cr -\sin \theta & \cos \theta}, 
\end{eqnarray} 
where $\mu = \mu_R + i \mu_I$, and 
\begin{equation}
\pmatrix{\chi_1 \cr \chi_2} = \pmatrix{\cos \theta & \sin \theta 
\cr -\sin \theta & \cos \theta} \pmatrix{\zeta_R \cr \zeta_I}.
\end{equation}
Hence $\zeta_R^2 + \zeta_I^2 = \chi_1^2 + \chi_2^2$, whereas 
$\zeta_R^2 - \zeta_I^2 = \cos 2 \theta (\chi_1^2 - \chi_2^2) - 2 \sin 2 \theta 
\chi_1 \chi_2$, and $2 \zeta_R \zeta_I = \sin 2 \theta (\chi_1^2 - \chi_2^2) 
+ 2 \cos 2 \theta \chi_1 \chi_2$.

\noindent \underline{\it Dark matter self-interactions}~:\\
The relevant trilinear couplings involving the physical $\chi_{1,2}$ 
dark-matter mass eigenstates are
\begin{eqnarray}
{\cal L}_3 &=& \sqrt{2} \lambda_{01} v h (\chi_1^2 + \chi_2^2) + 
\sqrt{2} \lambda_{12} u \eta_R (\chi_1^2 + \chi_2^2) \\ 
&+& {\eta_R \over \sqrt{2}} [ \mu_R \cos 2 \theta (\chi_1^2 - \chi_2^2) 
- 2 \mu_R \sin 2 \theta \chi_1 \chi_2 - \mu_I \sin 2 \theta (\chi_1^2 - 
\chi_2^2) - 2 \mu_I \cos 2 \theta \chi_1 \chi_2] \nonumber \\  
&+& {\eta_I \over \sqrt{2}} [ \mu_I \cos 2 \theta (\chi_1^2 - \chi_2^2) 
- 2 \mu_I \sin 2 \theta \chi_1 \chi_2 + \mu_R \sin 2 \theta (\chi_1^2 - 
\chi_2^2) + 2 \mu_R \cos 2 \theta \chi_1 \chi_2].  \nonumber
\end{eqnarray}
Let $m_{\chi_1} < m_{\chi_2}$, then $\chi_1$ is dark matter and interacts 
with itself through $h, \eta_R$, and $\eta_I$.  In particular, $\eta_I$ 
may be naturally light ($m_4 << v,u$), say 35 MeV, and be an excellent 
candidate for solving the cusp-core problem~\cite{kty16}.  
\begin{figure}[htb]
\vspace*{-4cm}
\hspace*{-3cm}
\includegraphics[scale=1.0]{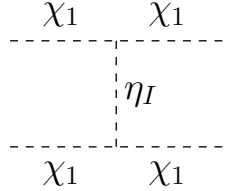}
\vspace*{-21.5cm}
\caption{Dark matter $\chi_1$ scattering by exchanging $\eta_I$.}
\end{figure}
As shown in Fig.~1, the elastic scattering cross section of $\chi_1$ by 
\begin{figure}[htb]
\vspace*{-4cm}
\hspace*{-3cm}
\includegraphics[scale=1.0]{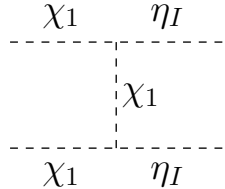}
\vspace*{-21.5cm}
\caption{Dark matter $\chi_1 \chi_1$ annihilation to $\eta_I \eta_I$.}
\end{figure}
exchanging $\eta_I$ is proportional to $m_{\eta_I}^{-4}$, whereas the 
annihilation cross section (as shown in Fig.~2) of $\chi_1 \chi_1$ to 
$\eta_I \eta_I$ (and $hh$, $\eta_R \eta_R$ if kinematically allowed) is 
proportional to $m_{\chi_1}^{-4}$.  Actually there is also the quartic coupling 
$\lambda_{12}$ which has been assumed negligible here for simplicity.  
To solve the cusp-core discrepancy, the condition is then roughly 
\begin{equation}
\left( {m_{\chi_1} \over m_{\eta_I}} \right)^4 \sim 10^{12} \left( 
{m_{\chi_1} \over {\rm GeV}} \right).
\end{equation}
This may be satisfied with $m_{\chi_1} \sim 100$ GeV, and $m_{\eta_I} \sim 
35$ MeV for example.

\noindent \underline{\it Linkage to the standard model}~:\\
There are three linkages between the new particles and the standard model, 
all coming from the SM Higgs boson $h$.
\begin{itemize}

\item The dark scalars $\chi_{1,2}$ are odd under $(-1)^D$.  They cannot mix 
with $h$, but they do interact through their trilinear couplings 
$\sqrt{2} \lambda_{01} v h (\chi_1^2 + \chi_2^2)$ as shown in Eq.~(9). 
\begin{figure}[htb]
\vspace*{-3.5cm}
\hspace*{-3cm}
\includegraphics[scale=1.0]{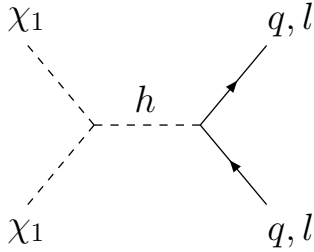}
\vspace*{-21.5cm}
\caption{Dark matter $\chi_1 \chi_1$ annihilation to SM particles through $h$.}
\end{figure}
This means that $\chi_1 \chi_1$ annihilation through $h$ to SM particles 
is possible as shown in Fig.~3 for relic abundance,
together with $\chi_1$ elastic scattering off nuclei as shown in Fig.~4 for 
its direct detection in underground experiments.
\begin{figure}[htb]
\vspace*{-4.5cm}
\hspace*{-3cm}
\includegraphics[scale=1.0]{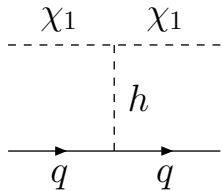}
\vspace*{-21.5cm}
\caption{Dark matter $\chi_1 $ elastic scattering off nuclei.}
\end{figure}
From the severe LUX limit~\cite{lux16} on direct detection, 
$\lambda_{01} < 0.01$ is implied~\cite{cskw13,fpu15,km15} which in turn 
gives much too small an annihilation cross section from Fig.~3 for obtaining 
the correct relic abundance unless there is a resonance effect, i.e. 
$m_{\chi_1}$ just below $m_h/2$.  However, there is also 
Fig.~2 in this model, which involves a different coupling, i.e. 
$(\mu_I \cos 2 \theta + \mu_R \sin 2 \theta)/\sqrt{2}$, 
thus evading this stringent constraint without any difficulty.

\item The heavy particle $\eta_R$ mixes with $h$ as shown in Eq.~(6). 
It will decay to SM particles through $h$.

\item The light particle $\eta_I$ does not mix with $h$ at tree level, 
but does so in one loop as shown in Fig.~5.  Note that if $\zeta$ is 
replaced by a Majorana fermion, then this mixing is forbidden, because 
$\eta_I$ would be odd under the $\gamma_5 \to -\gamma_5$ tranformation, 
whereas $h$ and $\eta_R$ are even.  Since most models assume that dark 
matter is a fermion~\cite{bdmsw14,imn14,adp15}, this mechanism is not 
applicable in those cases. 
For a stable light $\eta_I$, it could only annihilate to $e^-e^+$ through $h$, 
but then its cross section would be so small that it would overclose the 
Universe.
\begin{figure}[htb]
\vspace*{-4cm}
\hspace*{-3cm}
\includegraphics[scale=1.0]{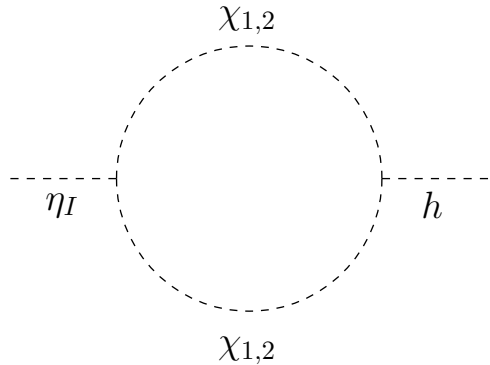}
\vspace*{-19.5cm}
\caption{One-loop finite mixing of $\eta_I$ with $h$.}
\end{figure}
The phenomenon of radiative Higgs mixing has only been discovered 
recently~\cite{m16}, and the effective quadratic $\eta_I h$ term is 
easily calculated to be
\begin{eqnarray}
m^2_{\eta_i h} &=& 2i \lambda_{01} v (\mu_I \cos 2 \theta + \mu_R \sin 2 \theta) 
\int {d^4 k \over (2\pi)^4} \left[ {1 \over (k^2 - m^2_{\chi_1})^2} - 
{1 \over (k^2 - m^2_{\chi_2})^2} \right] \nonumber \\ 
&=& {\lambda_{01} v  
(\mu_I \cos 2 \theta + \mu_R \sin 2 \theta) \over 8 \pi^2} \ln {m^2_{\chi_1} 
\over m^2_{\chi_2}}.
\end{eqnarray}
Note that $\mu_I \neq 0$ is crucial in obtaining this finite result.  If 
$\mu_I = 0$, then $\theta = 0$ also [see Eq.~(7)].  This would be the 
case if the dimension-two $m^2_3$ term were absent, because then $\mu$ may 
always be redefined as real.   The residual symmetry with $m_3^2 = 0$ 
would then become $Z_2 \times Z_2$, with $\eta_R \sim (+,+)$, $\eta_I 
\sim (+,-)$, $\zeta_R \sim (-,+)$, $\zeta_I \sim (-,-)$.  There would then 
be at least two stable dark-matter particles, say $\zeta_I$ and $\eta_I$. 
It is an interesting model in its own right, but not the subject of this 
paper.   With only the $Z_2$ residual symmetry considered here,  $\eta_I$ 
is not stable.  In order not to disturb the success of big bang 
nucleosynthesis, its lifetime should be less than about 1 s~\cite{z15}.  For 
$m_{\eta_I} = 35$ MeV so that it decays mainly to $e^- e^+$, and 
$m_h = 125$ GeV, this translates to
\begin{equation}
|\lambda_{01} (\mu_I \cos 2 \theta + \mu_R \sin 2 \theta) \ln (m_{\chi_1}^2 
/m_{\chi_2}^2)| > 0.05~{\rm GeV}.
\end{equation}
Since $\lambda_{01} < 0.01$ from direct detection, this requires 
$\mu_{R,I}$ to be greater than about 5 GeV.   For comparison, the 
annihilation cross section of about 1 pb (suitable for the correct 
relic abundance) is obtained for a value of about 20 GeV.   Note that 
$\eta_I$ also mixes radiatively with $\eta_R$, with mixing proportional 
to $(\mu_I \cos 2 \theta + \mu_R \sin 2 \theta)(\mu_R \cos 2 \theta - 
\mu_I \sin 2 \theta)$.  Note also that unlike most other proposals of 
a light mediator~\cite{lmz15,dky15,ghlw16}, $\eta_I$ does not contribute 
to the direct detection of 
dark matter, i.e. $\chi_1$, which is dominated here by $h$ exchange as 
shown in Fig.~4.

\end{itemize}

\noindent \underline{\it Some numerical examples}~:\\
Let $\mu_{eff} = \mu_I \cos 2 \theta + \mu_R \sin 2 \theta$, then the 
cross section for $\chi_1 \chi_1 \to \eta_I \eta_I$ (see Fig.~2) 
$\times$ their relative velocity is given by
\begin{equation}
\sigma \times v_{rel} = {\mu_{eff}^2 \over 16 \pi m^6_{\chi_1}}.
\end{equation}
Setting this equal to $3 \times 10^{-26} ~{\rm cm}^3/{\rm s}$ for the 
correct dark-matter relic abundance, a value of $\mu_{eff} = 19$ GeV 
is obtained for $m_{\chi_1} = 100$ GeV.

For the elastic self-scattering of $\chi_1$ through $\eta_I$ exchange 
(see Fig.~1), the cross section is given by
\begin{equation}
\sigma = {\mu_{eff}^4 \over 4 \pi m^4_{\eta_I} m^2_{\chi_1}}.
\end{equation}
For the benchmark value of $\sigma/m_{\chi_1} = 1~{\rm cm}^2/{\rm g}$ in 
self-interacting dark matter, 
a value of $m_{\eta_I} = 39$ MeV is obtained, using $m_{\chi_1} = 100$ GeV 
and $\mu_{eff} = 19$ GeV as before.

For the decay of the light scalar mediator $\eta_I$ to $e^-e^+$, its rate 
is given by
\begin{equation}
\Gamma = {m_{\eta_I} m_e^2 \over 16 \pi} \left[ {\lambda_{01} \mu_{eff} 
\over 8 \pi^2 m_h^2} \ln {m^2_{\chi_1} \over m^2_{\chi_2}} \right]^2.
\end{equation}
Using $m_{\eta_I} = 39$ MeV, $m_e = 0.511$ MeV, $m_h = 125$ GeV, $\mu_{eff} = 
19$ GeV, $m_{\chi_1} = 100$ GeV, $m_{\chi_2} = 200$ GeV, and $\lambda_{01} = 0.01$, 
the decay lifetime $\Gamma^{-1} = 0.07$ s is obtained.  This is short enough 
so that big bang nucleosynthesis may proceed without being disturbed.  

\noindent \underline{\it Production of the light pseudoscalar mediator}~:\\
The decay rate of the SM Higgs boson $h$ to a pair of $\eta_I$ is given by
\begin{equation}
\Gamma (h \to \eta_I \eta_I) = {\lambda_{02}^2 v^2 \over 4 \pi m_h}.
\end{equation}
Compared to the decay rate of $h \to \tau^- \tau^+$, i.e. 
\begin{equation}
\Gamma (h \to \tau^- \tau^+) = {m_h m_\tau^2 \over 16 \pi v^2},
\end{equation}
the two are equal if $\lambda_{02} = 3.7 \times 10^{-3}$.  Hence the 
decay $h \to \eta_I \eta_I$ may occur readily.  However, the lifetime 
of $\eta_I$, i.e. 0.07 s, is far too long for its decay product $e^- e^+$ 
to be observed within the Large Hadron Collider.   As for production 
by annihilation of dark matter at present, Sommerfeld enhancement may 
be possible~\cite{ls09,fky10}, in which case $e^- e^+$ production 
from $\eta_I$ decay may be observed.  On the other hand, this does not 
affect the fluctuations of the cosmic microwave background~\cite{s16} 
because $\chi_1$ is assumed to be significantly heavier than 10 GeV.

\noindent \underline{\it Discussion and synopsis}~:\\
Because the only connection between the dark sector and the standard model 
is through the one Higgs boson $h$, this model belongs to a general class 
considered in Ref.~\cite{prv08}.  However it is the first model which 
explains why a light mediator should occur, and why its mixing with $h$ 
is suppressed, both in terms of a symmetry and the details of how it is 
broken.  Note that in models of an $U(1)_D$ gauge boson with arbitrary kinetic 
mixing to the SM $U(1)_Y$, there is no fundamental understanding of why 
this mixing is so small.

To summarize, two complex scalars are introduced beyond the standard model.  
They are singlets of the SM, but transform under a dark $U(1)_D$ symmetry, 
with $\zeta \sim 1$ and $\eta \sim 2$.  The complete renormalizable Lagrangian 
containing them and the SM Higgs doublet is given in Eq.~(2).  The $U(1)_D$ 
symmetry is respected by all dimension-four and dimension-three terms, but 
are explicitly broken to $Z_2$ and $Z_4$ respectively by the dimension-two 
$\zeta^2 + \bar{\zeta}^2$ and $\eta^2 + \bar{\eta}^2$ terms.  In addition 
$\eta_R$ acquires a nonzero vacuum expectation value, so that the residual 
dark symmetry becomes $Z_2$.   Under this $Z_2$, the two mass eigenstates 
formed out of $\zeta_{R,I}$ are odd, the lighter one becoming dark matter, 
whereas $\eta_{R,I}$ are even, with $\eta_I$ much lighter naturally, 
corresponding to a would-be massless Goldstone boson from the spontaneous 
breaking of $U(1)_D$.  Hence $\eta_I$ acts as a naturally light mediator for 
the self-interacting dark matter.  Furthermore, it mixes radiatively with 
$h$ (a phenomenon discovered only recently) and decays fast enough 
to avoid disturbing big bang nucleosynthesis.  The correct relic 
abundance is obtained without conflicting with direct-search limits. 
This is thus a minimal model of self-interacting dark matter with 
all the desirable theoretical and phenomenological properties.

\noindent \underline{\it $U(1)_D$ as lepton number}~:\\
The global $U(1)_D$ considered in the above may be taken to be lepton number, 
under which neutrinos and charged leptons have $D=1$.  To connect $\zeta$ 
and $\eta$ to the SM leptons, the singlet right-handed neutrinos $N_R$ are 
added, with the allowed Yukawa interaction $\bar{\eta} N_R N_R$.  With the 
spontaneous and soft breaking of $U(1)_D$, $N_R$ acquires a large Majorana 
mass from $\langle \eta \rangle = u$ and neutrinos obtain small seesaw 
Majorana masses. The residual symmetry is $(-1)^D$, i.e. lepton parity, 
from which dark parity, i.e. $(-1)^D(-1)^{2j}$, may be derived~\cite{m15}.  
Hence $\zeta$ has odd dark parity, and all other particles have even 
dark parity.  The light mediator $\eta_I$ now decays also to two neutrinos 
through the $\bar{N}_R \nu_L \phi^0$ term.  This scenario is not as 
minimal, but it links the existence of neutrino mass to the dark sector, 
and offers a possible answer to the question: where does $U(1)_D$ come 
from? 

\noindent \underline{\it Acknowledgements}~:\\
This work is supported in part by the U.~S.~Department of Energy under 
Grant No.~DE-SC0008541.

\bibliographystyle{unsrt}

\end{document}